\documentclass[aip,jcp,reprint]{revtex4-1}
\usepackage{epsfig}
\usepackage{color}
\usepackage{amsmath}
\usepackage{amsfonts}
\usepackage{natbib}
\usepackage{notes2bib}

\newcommand{\avg}[1]{\left<#1\right>}
\newcommand{\len}[1]{\left|#1\right|}

\newcommand{\para}[1]{\left(#1\right)}

\newcommand{\kT}{\ensuremath{k_{\rm B}T}}

\begin{document}



\title{Quantum Tunneling Could Enable Proton Transfer Reactions on Titan}



\author{Henry W. Longo}
\author{Richard C. Remsing}
\email[]{rick.remsing@rutgers.edu}
\affiliation{Department of Chemistry and Chemical Biology, Rutgers University, Piscataway, NJ 08854}




\begin{abstract}
The surface of Titan, Saturn's largest moon, is rich in organics and is often suggested
to model early Earth environments. 
Titan's surface is cold, at a temperature of approximately 90~K, which prohibits most thermally activated
chemical reactions. 
However, quantum effects become more important at low temperatures and reactions that
are classically prohibited can often proceed through quantum mechanical pathways. 
Using path integral molecular dynamics simulations,
we investigate nuclear quantum effects on the thermodynamics
of model proton transfer reactions in liquid ethane.
We find that proton transfer can occur at Titan surface conditions through quantum tunneling.
Consequently, we estimate that nuclear quantum effects can enhance reaction rates by many orders of magnitude.
Our results suggest that nuclear quantum effects could facilitate prebiotic chemistry
on Titan, and quantum effects should be considered in future investigations.
\end{abstract}


\maketitle

\raggedbottom


\section{Introduction}
The Saturnian moon Titan has been likened to early Earth
because of its thick atmosphere, stable liquids on its surface, and a rich inventory of organic molecules~\cite{sagan1992titan,elachi2005cassini,lorenz2006sand,coustenis2007composition,paganelli2007titan,lorenz2008titan,niemann2010composition,mackenzie2021titan,magee2009inms,coustenis2008titan,raulin2012prebiotic,cable2012titan,lunine2020astrobiology}.
This begs the question, can Titan support prebiotic chemistry or even life? 
Photochemistry in Titan's atmosphere produces an assortment of molecules relevant to prebiotic chemistry~\cite{sagan1992titan,lorenz2008titan,waite2007process,raulin2012prebiotic,cable2012titan,horst2017titan,lunine2020astrobiology}. 
Many of these molecules can condense to the surface as solid cryominerals~\cite{vu2022Buta,cable2019co,maynard2016co,vu2014formation,maynard2018prospects,cable2018acetylene,cable2021titan,thakur2023molecular,thakur2024nuclear} or liquid hydrocarbons~\cite{mitri2007hydrocarbon,stofan2007lakes,cordier2009estimate,hayes2016lakes,mastrogiuseppe2019deep}. 
Liquid environments are generally more conducive to prebiotic chemistry,
but Titan's liquids are different from those that led to life on Earth;
Titan's lakes and seas are nonpolar liquid hydrocarbons at a frigid temperature of 94~K. 
These harsh conditions suggest that any prebiotic chemistry in Titan's cold hydrocarbon seas
must be different than that in terrestrial environments. 
Of particular concern is the low temperature on Titan's surface. 
Temperature-driven fluctuations that push a reaction over its thermodynamic barrier are exceedingly rare at cryogenic temperatures, such that thermally-driven chemical reactions are not expected in Titan's seas. 
As a result, any important prebiotic (or biotic) reactions must occur through other means. 
For example, a catalyst could reduce a reaction barrier far enough for thermal fluctuations to enable it to occur.
However, the thermal energy available on Titan's surface is approximately 0.2~kcal/mol, which is at least an order of magnitude smaller than barriers to simple reactions like proton transfer, and it is highly unlikely that a catalyst would reduce the reaction barrier this much. 
Alternately, non-thermal reaction paths may enable chemistry on Titan's surface. 
At low temperatures, the quantum mechanical nature of molecular systems becomes increasingly important,
and quantum mechanical reaction pathways can compete with or even dominate over classical pathways.
Of particular interest is quantum tunneling, where a quantum mechanical particle passes through an energy barrier instead of over it~\cite{mcmahon2003chemical,schreiner2020quantum,kaestner2014theory,schreiner2017tunneling,nitzan2006chemical}. 
Because of the low surface temperatures,
quantum tunneling through energy barriers could enable chemical reactions on Titan's surface
and play a key role in the development of alternative, non-aqueous biochemistries. 
%

\begin{figure*}[bt]
\begin{center}
\includegraphics[width=0.9\textwidth]{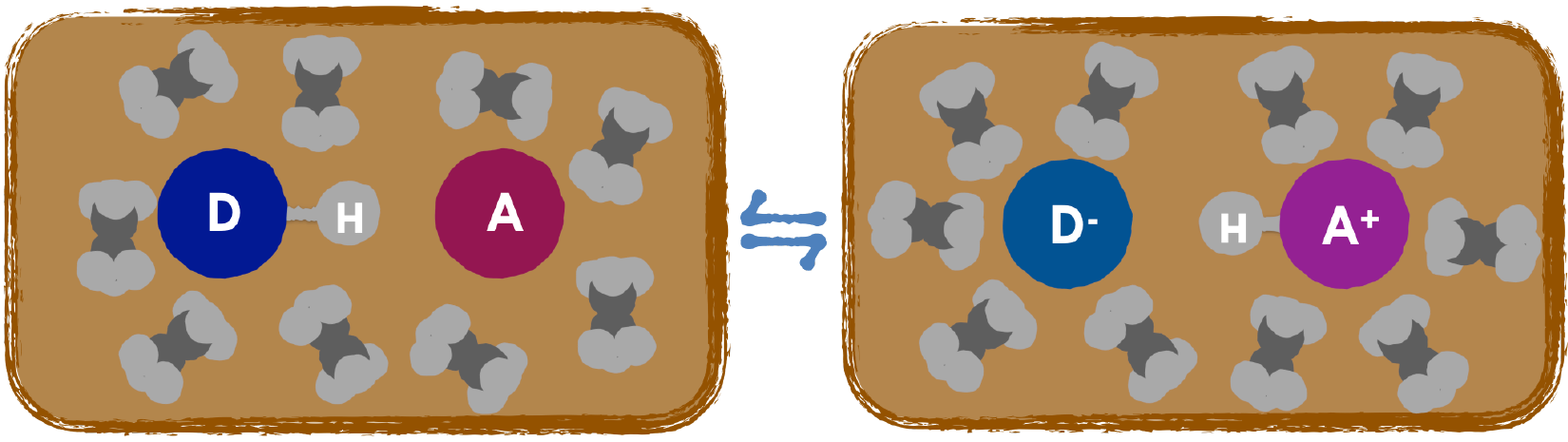}
\end{center}
\caption
{Sketch of the model proton transfer reaction studied in this work. 
A hydrogen-bonded donor-acceptor (D-A) complex is solvated by liquid ethane at 94~K. 
In the reactant state (left), the complex is composed of two neutral units.
In the product state (right), when the proton has transferred to the acceptor,
the donor and acceptor `molecules' are ionic.}
\label{fig:sketch}
\end{figure*}

%
Precise chemical reactions relevant to alternative biochemistries are not known, but
we can anticipate that fundamental processes like charge transfer are needed. 
Therefore, we focus on proton transfer reactions in Titan-like environments
and seek to determine if quantum tunneling can enhance the rate of these reactions at Titan's surface temperature. 
We study a simple model system that captures the essential physics of proton transfer~\cite{azzouz1993quantum,collepardo2008proton}. 
%
This model mimics a hydrogen-bonded donor-acceptor complex that transitions from neutral molecules
in the reactant state to an ionic complex in the product state, Fig.~\ref{fig:sketch},
and we model this proton transfer complex in liquid ethane as a model of Titan's seas. 
To account for the possibility of quantum tunneling, we model nuclear quantum effects
on reaction thermodynamics using path integral molecular dynamics simulations~\cite{FeynmanPathIntegrals,Chandler:JCP:1981,Berne:AnnRevPhysChem:1986,ceperley1995path,habershon2013ring}. 
Our results indicate that nuclear quantum effects effectively lower the free energy barrier for proton transfer
and as a result could increase reaction rates by tens of orders of magnitude. 
This finding suggests that nuclear quantum effects, such as quantum tunneling,
are a viable means for chemical reactions to occur on the surface of Titan,
opening avenues for investigation into non-aqueous prebiotic chemistry on cold worlds. 

\section{Results and Discussion}
\begin{figure}[tb]
\begin{center}
\includegraphics[width=0.49\textwidth]{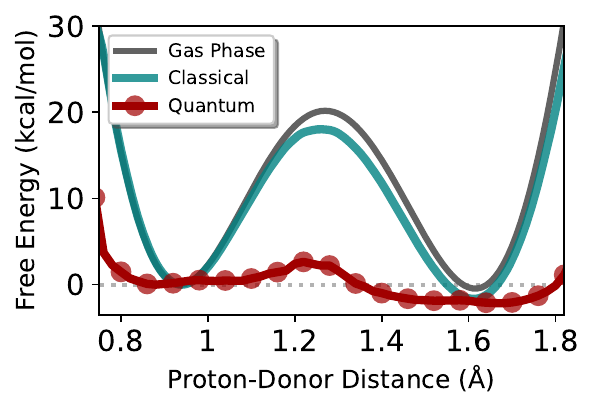}
\end{center}
\caption
{Free energy of proton transfer in gas phase and in liquid ethane at 94~K obtained with classical and quantum nuclei.}
\label{fig:fe}
\end{figure}

We focus on the model proton transfer reaction sketched in Fig.~\ref{fig:sketch}, 
which has a potential energy (gas phase free energy) that is nearly symmetric, Fig.~\ref{fig:fe}.
The reactant state consists of a neutral, dipolar, hydrogen-bonded donor-acceptor complex,
which is slightly more favorable than the product state.
In the product state, the proton has transferred from the donor to the acceptor, 
resulting in a negatively charged donor and a positively charged acceptor. 
Throughout the reaction, the donor-acceptor distance is fixed at $R=2.7$~\AA.
In a dielectric solvent, solvation stabilizes the product state due to solvation of the ionic complex, Fig.~\ref{fig:fe}.
To capture this stabilization through solvation, it is critical that the solvent model is able to capture the dielectric response of liquid ethane, and we use the recently-developed DC2 model of ethane to capture this dielectric response~\cite{thakur2021distributed}, which reproduces the experimental dielectric constant, $\varepsilon=1.94$.
The reduction of the free energy of the product state can be explained through the Bell model~\cite{bell1931electrostatic,zhao2020response},
\begin{equation}
\Delta G(\mu) = - \frac{1}{4\pi \epsilon_0} \frac{\varepsilon-1}{2\varepsilon+1}\frac{\mu^2}{R_S^3},
\label{eq:bell}
\end{equation}
where $\mu$ is the dipole moment of the donor-acceptor complex, $\epsilon_0$ is the vacuum permittivity, and $R_S$ is a parameter that represents an effective radius that results from approximating the complex as a sphere. 
The observed change in free energy of the ionic product state upon solvation, $\Delta G\approx-1.2$~kcal/mol, can be reproduced with a reasonable $R_S\approx7$~\AA, such that solvation indeed drives the changes in the free energy landscape. 
Moreover, the Bell model also predicts that when solvated, the product state has lower free energy; $\Delta G(\mu_{\rm P}) - \Delta G(\mu_{\rm R})<0$,
where $\mu_{\rm R}$ and $\mu_{\rm P}$ are the dipole moments of the reactant and product states. 
This consistency between the simple Bell model of solvation and our simulation results demonstrates that solvation by liquid ethane results in the observed changes in the free energy landscape between the `Gas Phase' and `Classical' solvated system shown in Fig.~\ref{fig:fe}. 
%

%
For light nuclei and cold temperatures, the quantum mechanical nature of nuclei can be important~\cite{markland2018nuclear}. 
We include a description of nuclear quantum effects on the proton transfer free energy
by using path integral molecular dynamics (PIMD) simulations (see Methods for details).
PIMD simulations leverage Feynman's rewriting of the partition function of a quantum mechanical particle
as the partition function for an isomorphic classical ring polymer with $P$ monomers or beads connected by harmonic bonds, whose spring constant is proportional to the particle's mass and temperature~\cite{FeynmanPathIntegrals,Chandler:JCP:1981,Berne:AnnRevPhysChem:1986,ceperley1995path,habershon2013ring}.  
One can then perform computer simulations of these classical ring polymers and accurately predict quantum mechanical averages to determine quantities like free energies and structural correlations. 
The spread of the ring polymer captures the dispersion of a quantum particle, and in the classical limit of large mass or high temperature, the beads of the ring polymer collapse to a single point. 
%

\begin{figure*}[tb]
\begin{center}
\includegraphics[width=0.85\textwidth]{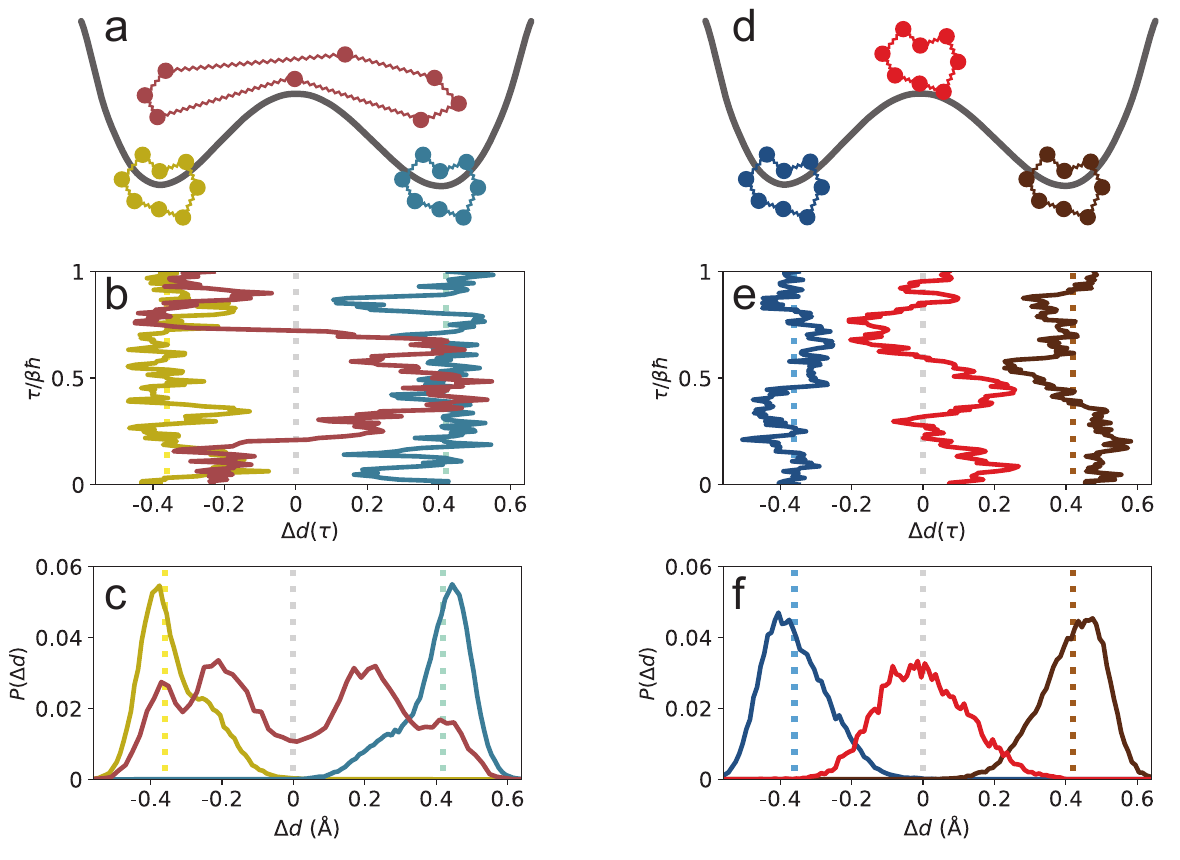}
\end{center}
\caption
{(a,d) Sketches schematically indicating configurations of the proton ring polymer when its centroid is the reactant minimum, transition state, and product minimum for (a) tunneling and (c) thermal, not tunneling, configuraitons.
(b,e) World-line diagrams showing $\Delta d$ as a function of ring polymer bead index (or imaginary time) for representative configurations of the centroid. Dashed line indicate the location of the reactant minimum, transition state, and product minimum.
(c,f) Probability distribution, $P(\Delta d)$, of the displacement of the proton away from the free energy barrier when the centroid is at the reactant minimum, the transition state, and the product minimum.
Panels correspond to (a-c) 94~K and (d-f) 300~K.}
\label{fig:world}
\end{figure*}

%
From our PIMD simulations, we find that nuclear quantum effects significantly alter the free energy landscape, Fig.~\ref{fig:fe}.
The quantum free energy landscape is broader than the classical one,
as is expected because quantum particles are `smeared out' when compared to their classical analogues, such that quantum particles can access classically forbidden regions. 
The quantum free energy barrier for proton transfer, $\Delta G^\ddag_{\rm q}\approx 2.8$~kcal/mol, is significantly lower than the classical barrier, $\Delta G^\ddag_{\rm c}\approx 18.0$~kcal/mol.
Importantly, with respect to the thermal energy, $\kT$, where $k_{\rm B}$ is Boltzmann's constant
and $T=94$~K is the temperature,
nuclear quantum effects lower the free energy barrier from approximately $97\kT$ in the classical system to
about $15\kT$ in the quantum system, greatly increasing the probability for the reaction to occur.
We can estimate the enhancement in the reaction rate due to quantum effects using transition state theory~\cite{Voth:JPC:1996,voth1989rigorous} as
\begin{equation}
\ln\para{\frac{k_{\rm q}}{k_{\rm c}}} \approx \frac{\Delta G^\ddag_{\rm c}-\Delta G^\ddag_{\rm q}}{\kT}\approx 82,
\end{equation}
or $\log_{10}(k_{\rm q}/k_{\rm c})\approx 35$,
such that nuclear quantum effects can enhance the rate of proton transfer by 35 orders of magnitude!
This enhancement strongly suggests that nuclear quantum effects can enhance chemistry on Titan's surface. 
We can assess if the proton is quantum tunneling through the free energy barrier with a world-line analysis of the proton ring polymer~\cite{ceperley1995path,miura1998ab,drechsel2014quantum}.
In this analysis, we examine the value of the reaction coordinate for each bead of the ring polymer in a configuration of the system. 
For ease of interpretation, we plot $\Delta d(\tau)$ as a function of the bead index (or imaginary time) $\tau$ in Fig.~\ref{fig:world}b,e,
where $\Delta d(\tau) = d(\tau)-d^\ddag$, $d^\ddag$ is the value of the proton-donor distance at the free energy barrier,
and $\tau/\beta\hbar$ normalizes the bead indices to span the range of zero to one.
To compare to a less quantum system, we also analyze proton ring polymers in a system that is simulated at the same fixed density but at a higher temperature of 300~K, where nuclear quantum effects should not be as significant.
We show three representative configurations of the proton at 94~K and 300~K and sketch what typical proton ring polymers look like for these three cases in Fig.~\ref{fig:world}.
Each configuration differs by the location of the center of mass or centroid of the proton ring polymer: one near the reactant minimum, one near the product minimum, and one near the transition state.
At both temperatures, $\Delta d(\tau)$ does not cross zero when the centroid is located in a free energy minimum, and the proton is not tunneling in those configurations. 
When the centroid is near the transition state, $\Delta d(\tau)$ obtained at 94~K displays two rapid crossings of zero (i.e. instantons) and spans the two minima, indicative of quantum tunneling in the system.
In contrast, $\Delta d(\tau)$ obtained at 300~K looks similar to the other curves and simply oscillates around the transition state, consistent with an absence of quantum tunneling at high temperatures.
%

\begin{figure}[tb]
\begin{center}
\includegraphics[width=0.49\textwidth]{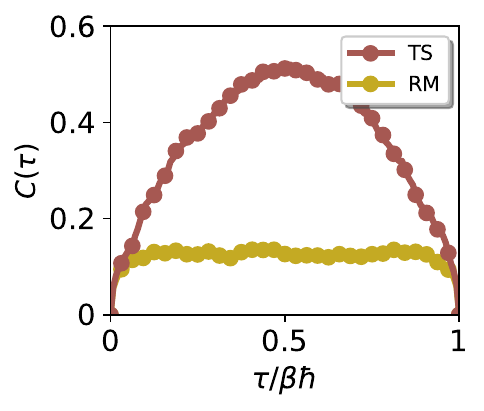}
\end{center}
\caption
{Correlation function, $C(\tau)$, of the proton-donor distance along the ring polymer, i.e. the imaginary time correlation function of $d(\tau)$,
for proton ring polymers at the transition state (TS) and in the reactant minimum (RM) for the quantum system at 94~K.}
\label{fig:ctau}
\end{figure}

%
We find further evidence for quantum tunneling at Titan conditions in the probability distribution of the relative proton-donor distance conditioned on the proton centroid position, $P(\Delta d | \Delta \bar{d})$, where $\Delta \bar{d} = P^{-1}\sum_{i=1}^P \Delta d_i$ is the value of $\Delta d$ for the centroid, obtained by averaging over the $P$ beads of the ring polymer, Fig.~\ref{fig:world}c,f.
When $\Delta \bar{d}$ is constrained to be at the reactant or product minimum, the proton is localized to that minimum and $P(\Delta d | \Delta \bar{d})$ exhibits a single peak.
At 94~K, when the centroid is at the transition state, $\Delta \bar{d}=0$, the distribution is bimodal.
The peaks of $P(\Delta d | 0)$ are centered at positive and negative values of $\Delta d$, consistent with quantum tunneling of the proton ring polymer.
At 300~K, when the system is more classical, $P(\Delta d | 0)$ is unimodal and centered at zero, consistent with the absence of quantum tunneling in proton transfer at this temperature. 
Finally, we also computed the imaginary time correlation function~\cite{drechsel2014quantum},
\begin{equation}
C(\tau) = \avg{\len{d(\tau)-d(0)}}^{1/2},
\end{equation}
which measures the correlation between values of the proton-donor distance along the ring polymer
from bead to 0 to bead $\tau$. 
When $\tau$ is normalized by $\beta\hbar$, it is restricted to values between 0 and 1,
and $C(\tau)$ is symmetric about $\tau/\beta\hbar = 1/2$, because of the cyclic nature of the ring polymer.
Within the reactant minimum, the ring polymer is localized, which is reflected
in the small plateau of $C(\tau)$ across most of the range of $\tau$.
In contrast, $C(\tau)$ is parabolic at the transition state at 94~K, signifying quantum delocalization of the proton
and providing further evidence of a quantum tunneling mechanism for proton transfer at Titan surface temperatures.

\section{Conclusion}
Our molecular simulations suggest that nuclear quantum effects can significantly lower barriers
to proton transfer in Titan-like environments, consequently enhancing reaction rates by tens of orders of magnitude. 
As a result, nuclear quantum effects could facilitate chemistry on Titan that is not possible through
classical, thermally-activated mechanisms. 
Biochemistry on Earth seems to leverage quantum effects when possible~\cite{markland2018nuclear,brookes2017quantum,mcfadden2018origins,cao2020quantum}, and one may anticipate that prebiotic or biotic processes on Titan's cold surface could exploit quantum effects as much as possible to enable chemical reactions on relevant timescales.
Our results also have implications for modeling Titan-like atmospheres. 
Atmospheric models informed by quantum chemical calculations of reaction barriers can produce first principles predictions of atmospheric compositions that are in reasonable agreement with Cassini observations~\cite{pearce2019consistent,pearce2020hcn}. 
Analyzing the resulting reaction networks can uncover key reactions in atmospheric production of specific molecules of interest.
However, many reactions in these networks involve light nuclei, and our findings suggest that quantum tunneling should be included when estimating rates of these reactions. 
While detailed path integral molecular dynamics calculations like those used here may be too expensive to perform for the more than one hundred reactions relevant just to HCN production, instanton methods appropriate to gas phase chemical reactions could be used to include tunneling of both light and heavy atoms in estimates of rate constants~\cite{claveau2023methane,richardson2018perspective,Beyer:2016aa,richardson2018ring}.
The results presented here are for a simple model
that captures the relevant physics involved in proton transfer reactions.
Consequently, we anticipate that the results are general and should be relevant to a wide array of proton transfer reactions.
But, quantification of nuclear quantum effects on reaction rates in specific molecular systems relevant to Titan is still needed. 
Similarly, we estimated reaction rates from free energy barriers using transition state theory, but these
estimates neglect dynamical recrossing effects~\cite{chandler1978statistical,voth1989rigorous}, and we leave the analysis of these dynamics to future work.
We also note that our focus here is on chemical reactivity on Titan, not the similarly important concern of solubility of reactants and products in liquid hydrocarbons.
However, nuclear quantum effects can significantly impact the dynamics of cryominerals~\cite{thakur2024nuclear}, and consequently could play a role in enabling solid-state chemistry on Titan. 
Importantly, we demonstrated that the effects of quantum tunneling on proton transfer in our model is small at ambient conditions on Earth in comparison to Titan temperatures. 
This finding is relevant to laboratory studies that mimic chemistry on Titan's surface by performing chemistry in hydrocarbons that are liquid at 300~K. 
While such work yields important insights into thermally-driven chemical reaction mechanisms and the role of the nonpolar solvent, the effects of quantum tunneling are implicitly neglected by studying systems at higher temperatures. 
Consequently, our conclusion that quantum tunneling may play an important role in chemistry on Titan's cold surface suggests that laboratory models need to be performed in the proper temperature range to ensure that quantum mechanical reaction mechanisms are allowed. 
This is generalizable to other cold worlds, and quantum mechanical effects like tunneling should become more important as the temperature is lowered. 
As a result, quantum tunneling may be relevant for chemical processes on cold planetary bodies in the solar system, such as Uranus~\cite{moses2020atmospheric} and Pluto~\cite{cruikshank2019prebiotic}, as well as cold exoplanets~\cite{Beaulieu:2006aa,molaverdikhani2019cold,hu2021photochemistry,Quick_2023}.

\section{Simulation Details}
To model proton transfer in liquid ethane, we adapt
the Azzouz-Borgis model for a hydrogen-bonded donor-acceptor complex~\cite{azzouz1993quantum}
and place this complex within a model of liquid ethane that reproduces its experimentally observed dielectric constant~\cite{thakur2021distributed}. 
The Azzouz-Borgis model represents the hydrogen-bonded complex with three sites, two of which serve as donor and acceptor molecules. 
The donor and acceptor sites are held at a fixed distance, and the third site, the proton, is allowed to move between the two while the overall complex is constrained to be linear. 
The charges on each site in the complex are dependent on the proton-donor distance, such that each component of the complex is neutral when the proton is covalently bonded to the donor
and the donor and acceptor become charged when the proton is covalently bonded to the acceptor. 
We modified the usual parameters of the Azzouz-Borgis model to better illustrate the impact of nuclear quantum effects on proton transfer in Titan conditions.
We tuned the parameters of the model to obtain a near symmetric free energy profile in gas phase that becomes more symmetric upon immersion in a solvent. 
Following the notation of Collepardo-Guevara, Craig, and Manolopoulos~\cite{collepardo2008proton},
the parameters of our modified model are
$a=11.2$~\AA$^{-1}$, $b=7.1\times10^{13}$~kcal/mol, $D_{\rm A}=112$~kcal/mol, $D_{\rm B}=115$~kcal/mol, $d_{\rm A}=0.90$~\AA, $d_{\rm B}=1.05$~\AA, $n_{\rm A}=n_{\rm B}=15$~\AA$^{-1}$, and $r_0=1.26927$~\AA. 
All other parameters are the same as those in the original model~\cite{azzouz1993quantum,collepardo2008proton}.
The results shown here are for a donor-acceptor distance of $R=2.7$~\AA, but qualitatively similar results were also obtained with $R=2.6$~\AA.
Proton transfer rarely occurs spontaneously during our simulations because of the high reaction barrier. 
To enable adequate sampling of the reaction, we use the well-tempered metadynamics (WTM) enhanced sampling approach to efficiently sample configurations of the system relevant to proton transfer. 
Our path integral molecular dynamics (PIMD) simulations with WTM used the LAMMPS software package~\cite{thompson2022lammps} interfaced with i-PI version 2 for path integral functionality~\cite{kapil2019pi} and PLUMED to include enhanced sampling~\cite{PLUMED}. 
The LAMMPS package was modified to include the interaction potentials in the Azzouz-Borgis model, including
the change in site charges with proton-donor distance, and the modified code is publicly available at github.com/remsing-group/titantunneling.
Our WTM-PIMD simulations were performed using 128 beads at $T=94$~K and 32 beads at $T=300$~K,
where the temperature was maintained using a PILE-g thermostat~\cite{ceriotti2010efficient}
and the equations of motion were integrated using the Cayley integrator with a timestep of 0.25~fs~\cite{korol2019cayley}. 
Simulations included 125 ethane molecules and one H-bonded donor-acceptor complex in a cubic box of length $L=22.53$~\AA. 
All nuclei were modeled quantum mechanically in the PIMD simulations. 
Converged free energy surfaces were obtained from trajectories of approximately 900~ps.  
The WTM simulations used a bias factor of 28 and a hill width and height of 0.018~\AA \ and 0.3~kcal/mol
to bias the proton centroid-donor centroid distance. 
Free energies as a function of the bead proton-donor distance were obtained by appropriate reweighting using PLUMED. 
%

\begin{acknowledgements}
This work is supported by the
National Aeronautics and Space Administration under grant number 80NSSC20K0609 issued through the NASA Exobiology Program.
\end{acknowledgements}

\bibliography{Titan}

\end{document}